# Does Mesophase $Ag_{4+x}Pb_2O_{6-z}$ ($0 \leq x \leq 1$, $0,5 \leq z \leq 0,75$) Appeal for a Point Contact $AT_c$ Superconductivity?

D. Djurek
*Alessandro Volta Applied Ceramics (AVAC)*, Kesten brijeg 5, Remete, 10 000 Zagreb, Croatia

**Abstract**

Weakly compacted fine particles of $Ag_{4+x}Pb_2O_{6-z}$ which is iso-structural with Byström-Evers oxide $Ag_5Pb_2O_6$ exhibit a novel class of physical properties, which includes an appearance of diamagnetism at t < 78 °C, with corresponding absence of electric resistance, both events sounding for possible superconductivity. Weak compaction was performed by controlled pressing of introductory components $Ag_2O$ and $PbO_2$. Pellets were heated from room temperature in an evacuated chamber, and solid state reaction proceeded at 340-350 °C in an oxygen atmosphere released by reaction itself (60 < p < 100 mbar). Electric and magnetic properties appealing for an ambient temperature superconductivity may be attributed to scattering processes indicated by tunnelling of Ag-Ag clusters between states localized on the point contacts formed between adjacent particles. As compared to the work on the subject during past two decades, reproducibility of output data is satisfactory, and their correlations to the input preparation conditions appear as to be conclusive.

# Introduction

During past two decades several indications of possible superconductivity (SC) in Pb-Ag-Cu-Sr-O [1], Pb-Ag-Cu-O [2] and $PbCO_3 2PbO + Ag_2O$ [3] systems at ambient temperatures ( $AT_c$) were reported. An analysis by x - ray diffraction recorded after solid state reaction revealed presence of the sub-valent compound $Ag_5Pb_2O_6$, as a single constituent which was metallic in the fused mixture, exhibiting a comparatively low electrical resistivity ($< 10^{-5}$ ohm-cm). $Ag_5Pb_2O_6$ has been synthesized by A. Byström and L. Evers [4], and assigned structure was reported as to be trigonal, with two Ag(1) cations positioned in z-axis channels. In subsequent work Jansen and co-workers reported [5] trigonal P(-)31m symmetry space group, and unit cell ( $Z = 3$) has dimensions a = 593.24 pm and c = 641.05 pm (Figure 1a,b). Two Ag(1) cations form the chains with alternating Ag(1)-Ag(1) distances $\delta_1$ = 309.3, and $\delta_2$ = 331.7 pm. Such a metallophilic polimerisation is promoted by the relativistic effect [6] which contributes to $d^{10} - d^{10}$ dimerisation between two Ag(1) cations. Formally, it is obvious that unit cell charge is unbalanced in $Ag_5Pb_2O_6$, and a rather heuristic model presented in this paper presumes that Ag(1) – Ag(1) dimers (clusters) provide only one valency to the surrounding lattice. Like in $Au_2$ clusters [7] 0,5 of electronic charge per Ag is assumed to be mutually exchanged between corresponding $5s^1$ states.

Subsequent experiments confirmed the relevancy of BE compound for SC events. However, reproducibility and stability on temperature cycling were poor. Observed transition temperatures were extended from 105 to 348 K, and by cycling of temperature these values gradually decreased. Experiments have also revealed two types of samples, one type demonstrated diamagnetism, with resistive transitions, while other type showed only low resistance states, with no accompanying magnetic effects. In this paper we turn our attention to the first type of samples, which sound for a Meissner type superconductivity.

X-ray diffraction technique and thermogravimetric decomposition, recorded in positive samples, revealed the presence of unreacted $Ag_2O$, or even metallic Ag, and it was presumed that $AT_c$ events are favoured in samples $Ag_5Pb_2O_6$ having Ag and/or oxygen defects, the former being probably proceeded in the c-axis channel. Attempts to produce Ag(1) defects by annealing of BE oxide in low pressure $O_2$ atmosphere at temperatures extending up to 400 °C were partly successful, but the method contributed only little to the bare reproducibility, since the control of input compaction parameters and of annealing procedure was a tedious job. The loss of oxygen and of Ag from BE oxide was favoured in the weakly compacted pellets, while it was unsuccessful in powders, or in powders strongly pressed in pellet. Compaction which involves the control of inter particle distance presumed an understanding of a rather complex structure of defect BE oxide and introduced many practical difficulties.

An example of a single shot result [9] clearly illustrates critical demands on the compaction procedure. Powdered BE oxide was prepared from $PbO_2$ and $Ag_2O$ at 110 bar $O_2$ and 310 °C. Powder was then pressed into pellet 8 mm in diameter and 1 mm thick, while uniaxial compaction pressure was 85 bar. Electrical resistivity of the pellet at room temperature (RT) was $7.3·10^{-5}$ ohm-cm. Heating in an evacuated chamber (Figure 2a) up to 400 °C resulted in a sintering, and subsequent cooling down to $LN_2$ temperatures stressed metallic behaviour with no resistive transition. Tensometric decomposition revealed $Ag_5Pb_2O_6$, i.e. no oxygen loss. A quite different situation was indicated in the pellet fabricated from the same powder batch, but compacted with 3.5 bar uniaxial stress. Resistivity at RT was considerably higher ($3.98·10^{-4}$ ohm-cm) and heating up to 400 °C in an evacuated chamber showed temperature dependence of the resistivity as shown in the inset of Figure 2b. During the heating up to 400 °C pellet released 0.125 mole $O_2$, chamber was evacuated, and cooling to RT (Figure 2b) in dynamic vacuum revealed the resistive transition at 75 °C, with final resistivity at 25 °C less than $5·10^{-8}$ ohm-cm. Tensometric decomposition released 0.29 moles $O_2$, and x-ray diffraction indicated presence of drained Ag. Obviously, certain fraction of cations $Pb^{4+}$ should be partly reduced to $Pb^{2+}$. Further attempts to reproduce this result, starting from the same powder batch, failed, which indicates a critical dependence on the particle distance, as well as on an application of more sophisticated compaction methods.

In point of fact, even four types of defects may appear in $Ag_5Pb_2O_6$. Draining of the Ag(1) cations from the c-axis channel offers two possible outcomes; firstly, remaining Ag(1) cations are bound in clusters, and secondly clusters are partly, or completely, fissioned to monovalent Ag (1). Third type of defects is associated with the partial reduction of $Pb^{4+}$ cations to $Pb^{2+}$, and finally, fourth type concerns to the oxygen content which may be less than that given by the stoichiometric value $O_6$.

Nearly all samples showing $AT_c$ phenomena were weakly coagulated powders [10], while no $AT_c$ event was observed in strongly uniaxially pressed pellets. Unfortunately, a decade ago, these observations fitted little to existing theories of superconductivity, models or physical properties known in ceramic materials exhibiting electric conductivity.

However, in the meantime it seemed that the field of research of two level systems (TLS) may adequately match the properties in weakly coagulated powders. TLS theories and models date back to seventies. [11,12,13], and TLS systems, as well as point contact spectrometry [14,15] performed in nano-junctions revealed fundamental significance of loosely bounded contacts. In this program an attempt is favoured to find common details between TLS and experimental data recorded in weakly bounded powders of $Ag_{4+x}Pb_2O_{6-z}$ (hereafter referred as to be *mesophase*).

Further support to relevancy of weakly bounded particles and TLS is provided by the experimental experience collected in our laboratory during two past decades.

Appearance of $AT_c$ phenomena was in no correlation with some particular fusion temperature, and positive samples were prepared in temperature range 292-413 °C. The same holds for oxygen pressures at which fusion proceeded, and range of $O_2$ pressures extended from vacuum to 290 bar. In several samples, being the mixture of BE compound and non conducting oxides, notably $Pb_3O_4$, diamagnetic fraction of magnetization at 100 K was considerably higher than the volume fraction of BE compound in the mixture [3]. Obviously, conducting particles do not act as carriers of diamagnetism, but as a part of some percolation interparticle mechanism extending in the whole sample. Furthermore, x-ray diffractograms recorded on positive samples showed no significant difference when compared to those recorded in normal samples. In some samples low resistance and diamagnetic states were sensitive to the vibrations coming from the rotary vacuum pump or transformer supplying the electric power in the furnace.

Recent experimental work of other groups further supports an assumption that $AT_c$ events are not associated with crystal structure in the grains, but with interfacial mechanism on the grain boundaries. R. König and co-workers [16] observed SC transitions at $0.62 \leq T_c \leq 1.38$ mK in carefully compacted platinum grains. Platinum itself is not superconducting, and transition temperature of powdered samples depended on the compaction strength. I. Bozovic and co-workers reported [17] superconductivity in well defined bilayer geometry consisting of an insulator ($La_2CuO_4$) and a metal ($La_{1.55}Sr_{0.45}CuO_4$), both compounds being not superconducting. Reported superconducting transition temperatures were $T_c = 15$ K, or $T_c = 30$ K, and result depends on the layering sequence.

**Synthesis**

In the supposed case when partial fission of clusters in the c-axis channel is not considered, more general formula, describing mesophase, may be represented as

$$Ag^{+1/2}_{1+x}[Ag^{+1}\ Pb^{+4}_{2-y}Pb^{+2}_{y}]O^{2-}_{6-z} \qquad (1)$$

Unit cell neutrality requires $1-x + 4y = 4z$, and in this paper we shall consider $0 \leq x \leq 1$, and $y = 1/2$. Oxygen defect is then given by $0.5 \leq z \leq 0.75$. Other possibilities including $z = 1$, $y = 1$, associated also with partial fission of clusters, will be published in forthcoming papers.

The powders of $Ag_2O$ and $PbO_2$ (manufactured by Kemika, Zagreb) were mixed in a magnetic stirrer with ethanol as a mixing agent, and corresponding stoichiometric proportions were calculated according to the equation

$$(2 + x/2)Ag_2O + 2PbO_2 = Ag_{1+x}[Ag_3Pb_2O_{6-z}] + (x/4 + z/2)O_2 \qquad (2)$$

The loosely pressed pellets were fabricated in the small compaction system, as shown in Figure 3. Fine screw mechanism (1) drives the pressing stamp (2) made of plexi-glass and compaction of the powder is performed in a die made of four plexi-glass segments (4) which are fixed by plexi-glass ring (3). The die bottom consists of a board (4) with printed four contacts, which serve for the four probe measurement of the electric resistance during the compaction. Electric conductivity of the powdered mixture comes from $PbO_2$, and depending on the compaction force it may be reduced from $\sim 10^7$ ohms to the order of $10^{-2}$ ohm. Resistance of the compacted powder is very reliable input parameter for preparation of pellets prior to the heating and fusion, and it is properly correlated with measured physical properties after the fusion. Additional control of compaction is provided by the laser distance meter (L) which measures the relative displacement δ of the plates A and B. δ has a characteristic dependence on the compaction force provided by the calibrated elastic spring (6) and this dependence may be also correlated to the output physical data. During the compaction four gold wires 50 microns in diameter may be introduced in the pellet, which facilitates the subsequent measurements of the electric resistance.

The $Ag_2O/PbO_2$ mixture, being either powdered, or in the form of the pellet was embedded in the home made hot stage consisting of a reaction cell made of platinum, and the cell was part of the reaction chamber prepared in the form of the stainless steel tube 15 mm of inner diameter. The pressure in the chamber was measured by the use of absolute capacitance gauge with an accuracy of 0.4 mbar. The hot part of the tube positioned in the furnace was designed with an empty volume small as possible, in order to minimize the fraction of volume in which pressure is affected by the temperature. This fraction has been estimated by the use of 1 bar argon gas heated from RT up to 600 °C, when pressure was increased to 1.022 bar. An additional correction is related to the natural increase of the pressure in the reaction chamber, as a result of stainless steel out-gassing and porosity of the tube wall. This correction amounts 3 and 5 mbar by heating up to 400 and 600 °C respectively. The compound used for the calibration of the chamber volume was $PbCO_3$ preheated *in vacuo* for 24 hours at 100 °C, in order to remove the humidity, and it was released 362.7 mbar $CO_2$ per mole of $PbCO_3$, being measured after cooling from 400 to 24 °C. The evaluated volume of the reaction chamber was calculated $V = 68.05$ cm$^3$ from the gas equation.

Chamber containing the sample was evacuated at RT and heated up to 200 °C, in order to remove $CO_2$ from traces of $Ag_2CO_3$, and the mixture was further heated up to 340 – 360 °C, when reaction starts with corresponding release of $O_2$, in agreement with Eq. 2.

Depending on the compression force exerted to the pellet, three outputs are possible. Firstly, in an un-compacted powder mixture $Ag_2O/PbO_2$, $Ag_2O$ decomposes to $2Ag + 1/2O_2$ by heating at t > 220 °C, and fusion to the mesophase fails. The same happens for moderately high compaction pressures (p > 1.8 bar). For medium compaction pressures (0.1 – 1.5 bar) and $x = 0$, fusion of the mesophase at 340-360 °C proceeds successfully. The reaction usually lasts 1-2 hours for $x = 1$, and 24 hours for $x < 0.5$ Final pressure of the released oxygen $(x/4 + z/2)O_2$ for $x = 0$ is 86.4 mbar, measured at 24 °C and for starting pellet $m = 500$ mg. Since reaction fails for $O_2$ pressures < 20 mbar detailed diagram connecting compaction density and $O_2$ pressure for selected fusion temperatures should be evaluated in the subsequent work. Evaluated coexistence diagram for mesophase (MF) is shown in Figure 4, and densities belong to the values measured after fusion. It is evident a remarkable phenomenon that an outcome of chemical reaction depends on the compaction density, and fusion is successful only in a rather narrow range of the densities of compacted components. The mesophase, although prepared in comparatively low oxygen pressures (<100 mbar) is surprisingly pure, and outer contours of the coexistence diagram were estimated by the use of x-ray diffraction ($CuK_\alpha$), when elementary Ag and $PbO_2$ ($Pb_3O_4$) were included only in traces, with relative diffraction intensities $I_R < 1.5$ percents. Figure 5 shows structure refinement of BE oxide prepared at 190 bar $O_2$ and 350 °C, by the use of FullProf Program, and Figure 6 is diffractogram of the mesophase $Ag_{4.5}Pb_2O_{5.36}$ reground into powder. By x there is marked the strongest diffraction of $Pb_3O_4$ (d = 338 pm), and of relative intensity $I_R = 1.4$ percents, coming from the unreacted part of $PbO_2$. Corresponding unreacted fraction of $Ag_2O$ is visible in tensometric decomposition in temperature range 220-380 °C. Unreacted molar fraction in mesophase was estimated to be 0.3-0.5 percents. It should be noted that the crystal structure, oxygen and Ag content in the mesophase are not affected by a gentle reground procedure.

**Physical properties**

An additional evaluation of the oxygen defect $z$ is provided by decomposition of $Ag_{1+x}$ $[Ag_3Pb_2O_{6-z}]$ in the same chamber, and according to the equation

$$Ag_{1+x}[Ag_3Pb_2O_{6-z}] = (4+x)Ag + 2/3\, Pb_3O_4 + (5/3 - z/2)O_2 \qquad (3)$$

Subsequent decomposition of $Pb_3O_4$ is inhibited at 510 °C because of the presence of the pressure of released $(5/3 - z/2)\, O_2$. Removal of this oxygen by pumping must be careful, since $Pb_3O_4$ decomposes fast, and, in order to avoid an uncontrolled loss of the oxygen released from $Pb_3O_4$ during the manipulation of the vacuum valves cell was cooled down to 400 °C, oxygen pumped away, and by reheating up to 550 °C $2/3Pb_3O_4$ decomposes to $2PbO+1/3O_2$. Decomposition diagram is shown in Figure 7. The ratio of released moles from two decomposition stages provides an additional possibility to control the oxygen defect $z$, and for $x = 1$, $z = 0$ it should be 5 : 1.

The values of oxygen $z$ recorded for selected $x$ are given in Table 1. For the reasons of symmetry, it is evident that $Pb^{4+}$ should be partly ($y=½$) reduced to $Pb^{2+}$, and only clustered Ag(1) cations are possible in the c-axis channel. The fraction of clusters is $(1 + x)/2$.

The measurements of electric resistance during the fusion and cycling of temperature between RT and 400 °C was performed in the reaction chamber similar in size and performance to that described above, but supplied with sample holder and electric leads, as described in the previous work [18]. The reaction chamber was evacuated at RT by the rotary pump, and powdered mixture was fused at 355 °C for 24 hours. The pressure in the chamber coming from the released $O_2$ was 63 mbar, and after cooling to RT oxygen was pumped out, which is followed by an introduction of air. It was observed that heating and cooling in an oxygen atmosphere with pressures extending up to 200 bar affect the physical properties, but in this report, excluding data in Figure 2, only experiments performed in air are reviewed.

Resistance measurements in dc field were performed by the use of Keithley AC/DC 6621 current source, and voltage drop was recorded by Keithley 2700 DMM. The ac resistance measurements were not performed at temperatures higher than transition temperature. The reason is provided by the high sensitivity of output data on the frequency, which may be also recognized as a delayed voltage response (0.01 – 0.7 seconds) at t > 100 °C to the step-like increase of the current, probably as a result of high relaxation times in the point contact continuum.

The temperature dependences of the electrical resistances recorded for different introductory compaction resistances show a strong correlation to the compaction strengths of the powder mixture prior to the heating. Resistance gradually decreased with number of cycles, performed from RT to 380 °C and is stabilized after 3 - 5 cycles. Figure 8a, 8b and 8c show the temperature dependence of the resistance for respective compaction resistances $1.3 \cdot 10^6$, $1.7 \cdot 10^5$ and $2.8 \cdot 10^3$ ohm, and respective pellet thicknesses were 1.1, 0.91 and 1.3 mm. Low cooling rate (15 deg/hour) was ensured by linear temperature programmer. Resistances increase strongly with decreasing temperature, and sample with lowest density (2.3 g/cm$^3$) reaches maximum of $2.4 \cdot 10^3$ ohms near 80 °C, which is followed by decrease. By further cooling to liquid nitrogen temperatures resistances displayed in Figure 8a and 8b reach no zero value, and diamagnetism measured at RT was weak, ranging from $5 – 8 \cdot 10^{-4}$ emu/cm$^3$. Sample compacted up to density 3.11 g/cm$^3$ (compaction resistance ~ $2.8 \cdot 10^3$ ohm) exhibits resistivity (Figure 8c) less than $3 \cdot 10^{-8}$ ohm-cm at t < 5 °C, which was justified by SR 730 lock-in amplifier operating at 175 Hz and 30 mA. Critical current density at 12 °C was ~ 2.6 A/cm$^2$.

It is interesting that further increase of compaction pressure, with resistances ranging 300 – 1300 ohms result in series of samples with transition temperatures $T_c$ < 0°C, but these data are not reviewed in this paper.

Increase of the compaction density resulted in an increased quality of the resistive transitions, as it is shown in Figure 8 for $x = 0.66$ and compaction resistance 65 ohm.

Measurement of the ac magnetic susceptibility posed an unexpected difficulty. An application of even small exciting magnetic fields destroys mechanically the weakly compacted pellets, and they were pulverized, probably as a result of strong inter-grain forces coming from the high magnetization gradients. In order to overcome this difficulty powder was compacted by the use of the same compaction device as above, but compaction die was the ceramic tube with inner diameter 4 mm.

Such a die can also be used for a quick preparation of samples subsequently subjected to measurements of the diamagnetism at RT after fusion, and completed device is shown in the inset of Figure 7. Powdered mixture (1) of $Ag_2O/PbO_2$ is embedded in the tube (2) supported by a copper plate (3), serving also as one electrode for the resistance measurements. Other electrode is copper stud (4) insulated by a mantle (5). Powder is tapped by the use of finger force on the stud until desired resistance is indicated on the measuring instrument. After fusion at 355 °C and reheating to 375 °C ac susceptibility measurements were performed in a standard configuration consisting of the primary and two opposed secondary coils. A commercial YBCO sample shaped to the similar size was used as a standard. The data are shown in Figure 10, and diamagnetic fraction is nearly 1/3 of full diamagnetism $1/4\pi$ emu/cm$^3$, which is the highest value at 0 °C reached in these experiments. The broad transition and comparatively small diamagnetic fraction are probably due to the inhomogeneity coming from the pressure gradient across the sample ~ 4.2 mm in thickness. Despite cited difficulties diamagnetic transitions are convincing and fairly reproducible.

**Discussion**

A weak compaction and controlled inter-grain distance of mixtures $Ag_2O/PbO_2$ before solid state reaction, or BE oxide $Ag_5Pb_2O_6$ might result to the unique chemical, metallurgical and physical properties

The direction of the chemical reaction depends on the precise control of the compression force exerted to the mixture of components before reaction, and this strange phenomenon is amazing in the light of empirical stereotypes about the kinetics of chemical reactions. At low oxygen pressures (less than 100 mbar) and t < 400 °C BE oxide $Ag_5Pb_2O_6$ given in powdered or strongly compacted form releases no oxygen, while $O_2$ is released from a weakly compacted pellet, with simultaneous preservation of the original crystal lattice structure. Finally, physical properties, like appearance of diamagnetism and low resistance states, are critically dependent on the intergrain distance, and this was demonstrated in samples indicated in Figure 2, which exhibited a rather good reproducibility when it was going on the oxygen release, but opposite is true for the tests on the possible superconductivity.

The most intensive diamagnetic fractions were recorded in samples with $x=½$, but significant data were also obtained for other values of $x$, including $x=1$. However, samples are inhomogeneously strained, and it is worth emphasizing that coexistence diagram shown in Figure 4 might be split in several regions, which implies more careful control of strains. The common property to the samples exhibiting $AT_c$ phenomena is partial reduction of $Pb^{4+}$ cations, which may affect the ballistic and tunnelling properties of Ag – Ag clusters on the point contacts, and, according to tight binding models [19], clusters are suspended between two grains serving as semi-infinite current leads. Indeed, independent experiments [20] have shown that such species exist, and may be generated by ion bombardment of metal surfaces.

Point contact physics and accompanied two level energy properties on the contacts provide an ample space for speculations and models which may match the phenomena reported in this work. In addition, powder metallurgy methods used in these experiments offer some important advantages when compared to the single nanoconstrictions and spear anvil type contacts, currently in use. Point contacts performed by careful pressing of sub-micron particles exhibit the comparatively small contact area, and the size of the contact can virtually be comparable to the Fermi wavelength $1/k_F$. Furthermore, anvil type contacts are sensitive on the external contamination, adsorbates and oxidation. By heating of mesophase *in vacuo* up to 400 °C typical release of the $O_2$ amounts $10^{-6}$ moles, which approximately corresponds to unsignificant 0.2 $O_2$ molecule per grain. In addition, oxide BE compound is no subject of further oxidation and is stable against humidity.

To date, an explanation of physical properties listed in this paper by the superconductivity might be somewhat premature, since it is going on new classes of materials with properties beyond the standard knowledge and incompatible to common sense. In the last instance, superconductivity is understood as a macroscopic coherent quantum state, and such a state must have properly defined and stable underlying material structure. In the mesophase $Ag_{4+x}Pb_2O_{6-z}$ with strong intergrain interaction this structure involves fluctuating strains and deformations modulated over many grain diameters. Strains are strongly coupled to the passing dc current, and this was demonstrated by current pulses (100 – 200 mA) which can destroy the pellet and convert it to the pulverized state. Certainly, a classical sintering methods and stabilization of the structure by heating at high temperatures are not preferred in future work. More promising method is mixing of mesophase with other insulating oxides, program which is in the course.

**Conclusion**

It has been indicated that Byström-Evers oxide $Ag_5Pb_2O_6$ may be obtained in iso-structural form $Ag_{4+x}Pb_2O_{6-z}$ ($0 \leq x \leq 1$, $0 \leq z \leq 0.75$), and loosely coagulated grains of $Ag_{4+x}Pb_2O_{6-z}$ form mesophase with exclusive point contact properties.. It is supposed that highly mobile Ag - Ag clusters, positioned in the crystal lattice c-axis channel, tunnel between nearly degenerate point contact states on grain boundaries, which may result in different phases exhibiting in mesoscopic point contact continuum semiconducting, metallic or semimetallic properties. Some phases obey electric and magnetic properties characteristic for ambient temperature superconductivity with $T_c \sim 78$ °C, result being claimed during past two decades. In contrast to previous reports, when powders were strongly pressed in pellets, loosely coagulated mesophases exhibit reproducible $AT_c$ events.

Samples are brittle and exhibit poor stability on temperature cycling, but they are reviewed in this paper for the sake of physical clarity and scientifically conclusive claims. Next research program includes the fabrication of stable and practical samples by the alloying $Ag_{4+x}Pb_2O_{6-z}$ with insulating oxides having carefully tuned elastic and viscoelastic properties. Dilutions extend down to volume fraction of $Ag_{4+x}Pb_2O_{6-z}$ in oxides less than 0,10, which also implies an economic feasibility due to the considerably smaller silver consumption. Several additional iso-structural phases of $Ag_{4+x}Pb_2O_{6-z}$ will be reviewed in next papers.

**Acknowledgement**

Author is indebted to Dr M. Paljević and Professor A. Tonejc for cooperation during initial part of the project. Dr Z. Medunić visited 1992. *Laboratoire de Cristallographie & Centre de Recherche sur les tres Basses Temperature*, Grenoble, when BE oxide has been firstly recognized in fused Pb-Ag-O mixtures.

**Figure captions**

**Fig. 1** Trigonal unit cell of Byström-Evers oxide $Ag_5Pb_2O_6$ (a), and c-axis view (b).

**Fig. 2** Temperature dependence of the resistivity of BE oxide $Ag_5Pb_2O_6$; (a) strongly compacted powder (85 bar), (b) weakly compacted powder (3.5 bar). Inset shows heating cycle.

**Fig. 3** Mechanical device for a weak compaction of powdered mixtures $Ag_2O/PbO_2$; (1) driving screw, (2) pressing stamp, (3) outer die ring, (4) die segments, (5) printed board, (6) calibrated elastic spring. Laser distance meter serves for compaction control.

**Fig. 4** Coexistence diagram of the mesophase (MF) $Ag_{4+x}Pb_2O_{6-z}$ evaluated for various densities, measured after solid state reaction, and for various $x$.

**Fig. 5** X – ray diffractogram BE oxide $Ag_5Pb_2O_6$ and corresponding refinement of the structure, performed by the use of FullProf Program.

**Fig. 6** X – ray diffractogram of the gently reground mesophase $Ag_{4.5}Pb_2O_{5.36}$.

**Fig. 7** Tensometric decomposition of $Ag_{4.5}Pb_2O_{6-z}$; (a) up to remainder $2/3Pb_3O_4$, (b) decomposition of $2/3Pb_3O_4$. Inset shows a simplified device for compaction of the powdered mixtures $Ag_2O/PbO_2$ (1), embedded into ceramic tube (2), and supported by a copper plate (3) serving as a contact for resistance measurements. Other contact electrode is copper tapping stud (4).

**Fig. 8** Temperature dependence of the electric resistance of the samples $Ag_{4.5}Pb_2O_{5.36}$ fabricated from $Ag_2O/PbO_2$ powders, and compacted by various compaction forces indicated by corresponding resistances; (a) 1.3 Mohm, (b) 0.17 Mohm, (c) 2.8 kohm. Densities were measured after the solid state reaction.

**Fig. 9** Temperature dependence of he electric resistance of the mesophase $Ag_{4,66}Pb_2O_{5,42}$ indicated by compaction resistance 65 ohm.

**Fig. 10** Temperature dependence of ac susceptibility of the mesophase $Ag_{4.5}Pb_2O_{5.35}$ indicated by compaction resistance 145 ohm.

**Table caption**

**Table 1** Tensometric evaluation of the oxygen defects ($z$) and calculated reduction of $Pb^{4+}$ ($y$) for various fractions ($x$) in the mesophase $Ag_{1+x}[Ag_3Pb^{4+}_{2-y}Pb^{2+}_y]O_{6-z}$. Ag(1) cations are assumed to be completely clustered.

| $x$ | $y$ | $z$ |
|---|---|---|
| 0,00 | 0,501 | 0,76 ÷ 0,015 |
| 1/3 | 0.562 | 0,73 ÷ 0,03 |
| 1/2 | 0,515 | 0,64 ÷ 0,015 |
| 2/3 | 0,505 | 0,59 ÷ 0,035 |
| 1,00 | 0,480 | 0,48 ÷ 0,02 |

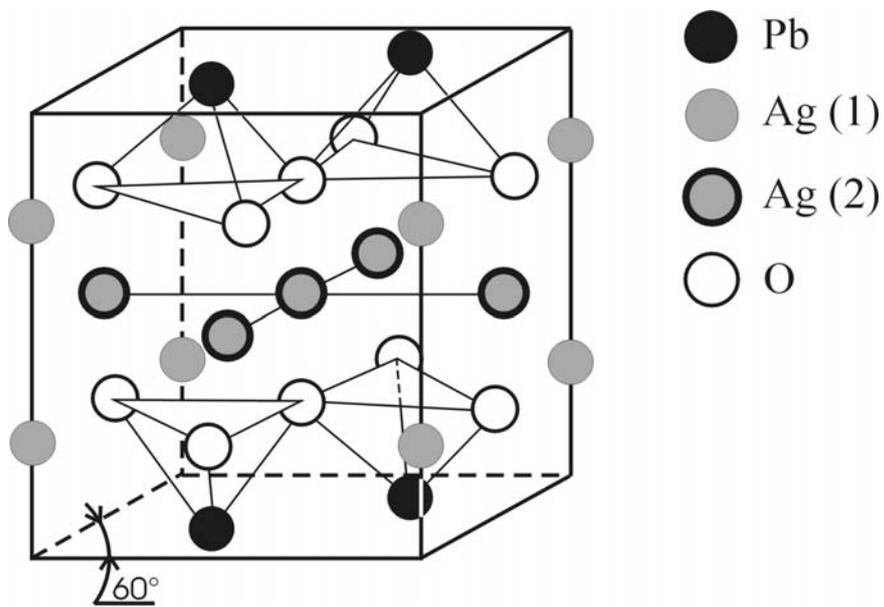

**Figure 1a**

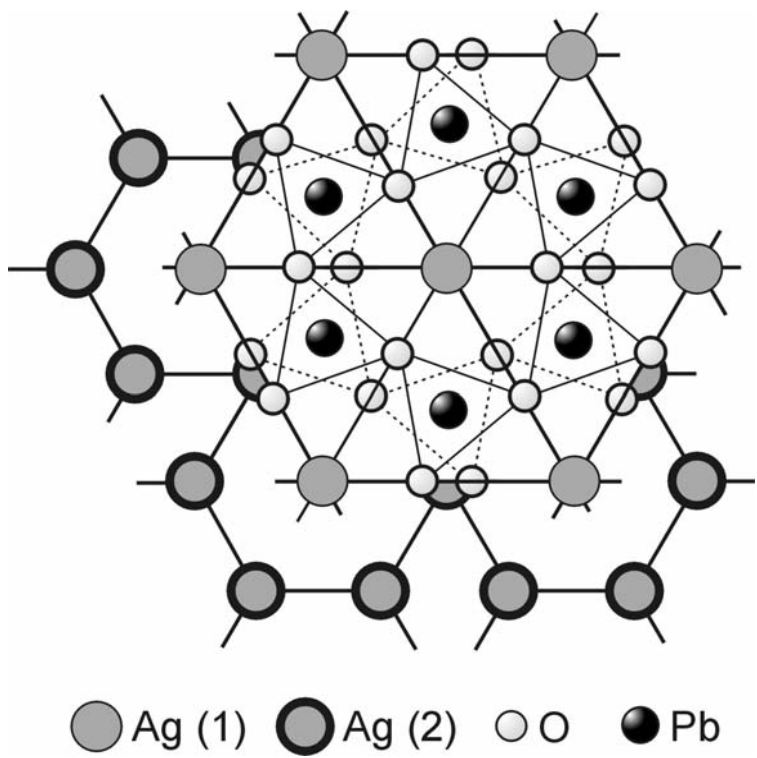

**Figure 1b**

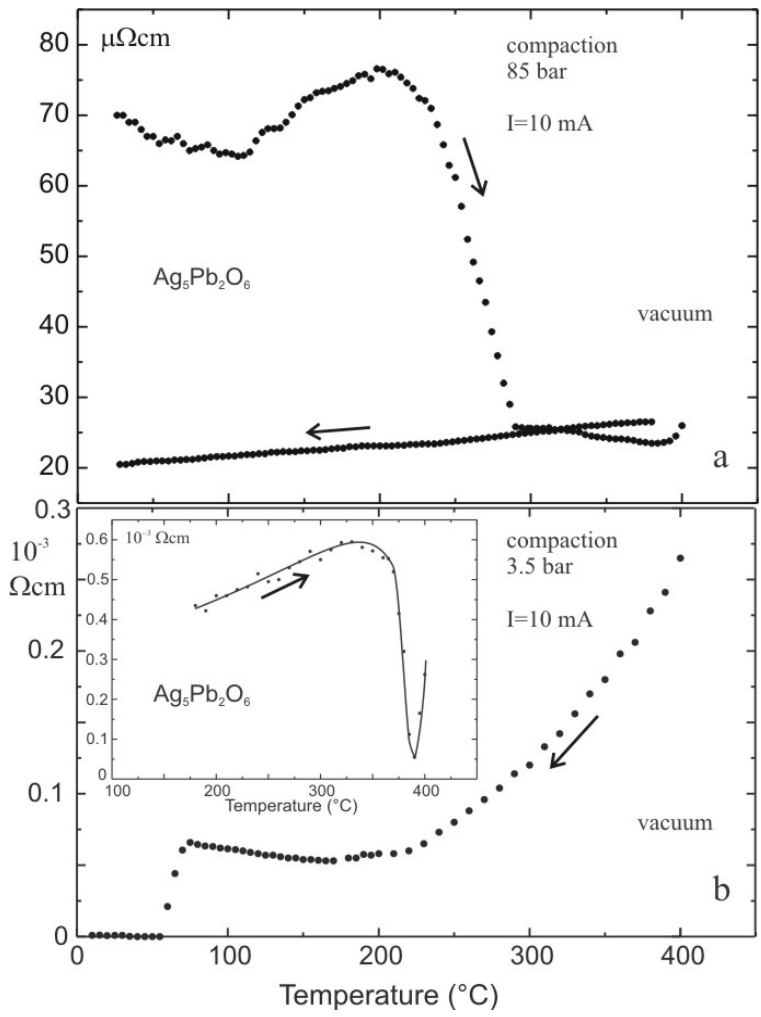

**Figure 2**

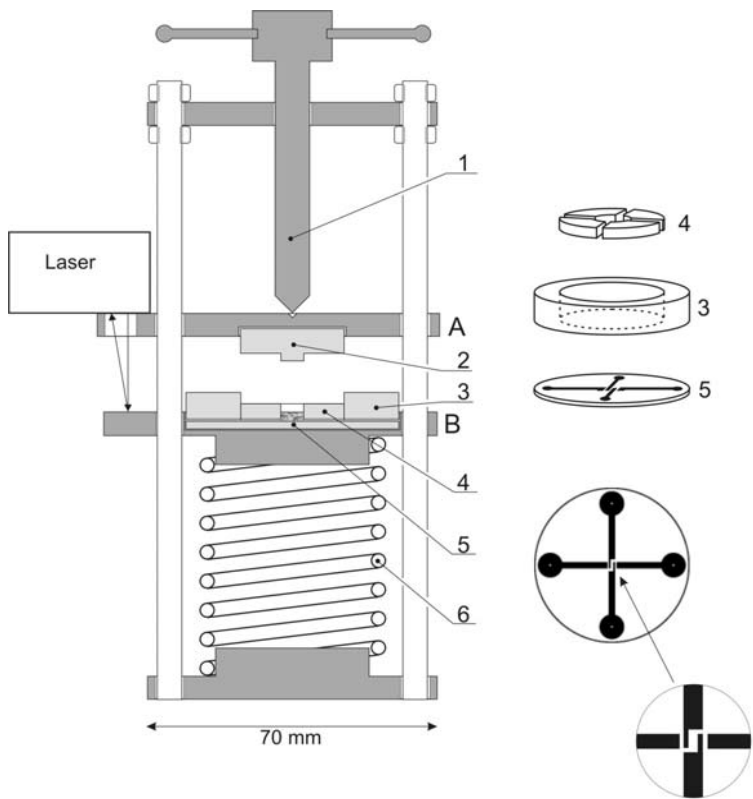

**Figure 3**

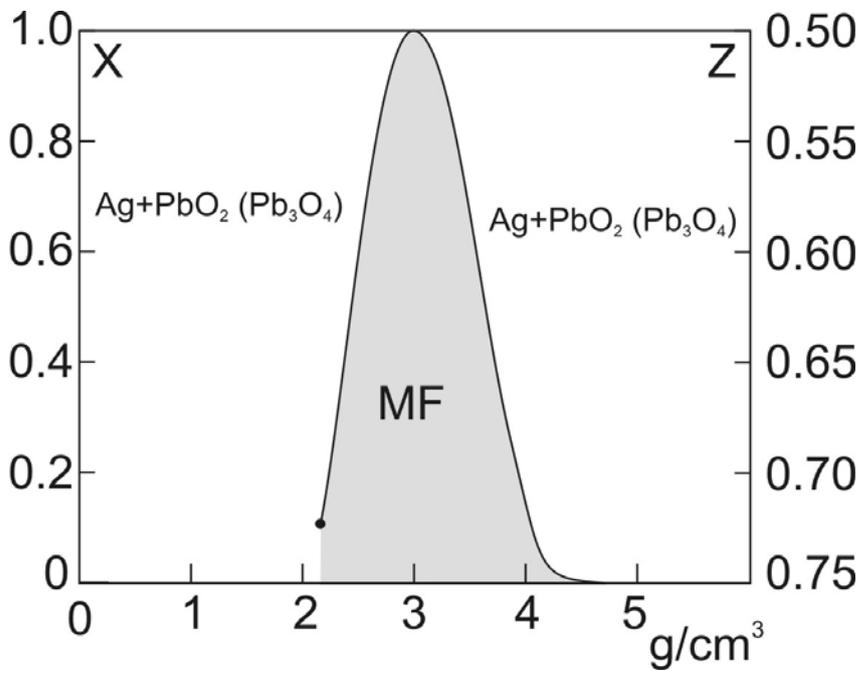

**Figure 4**

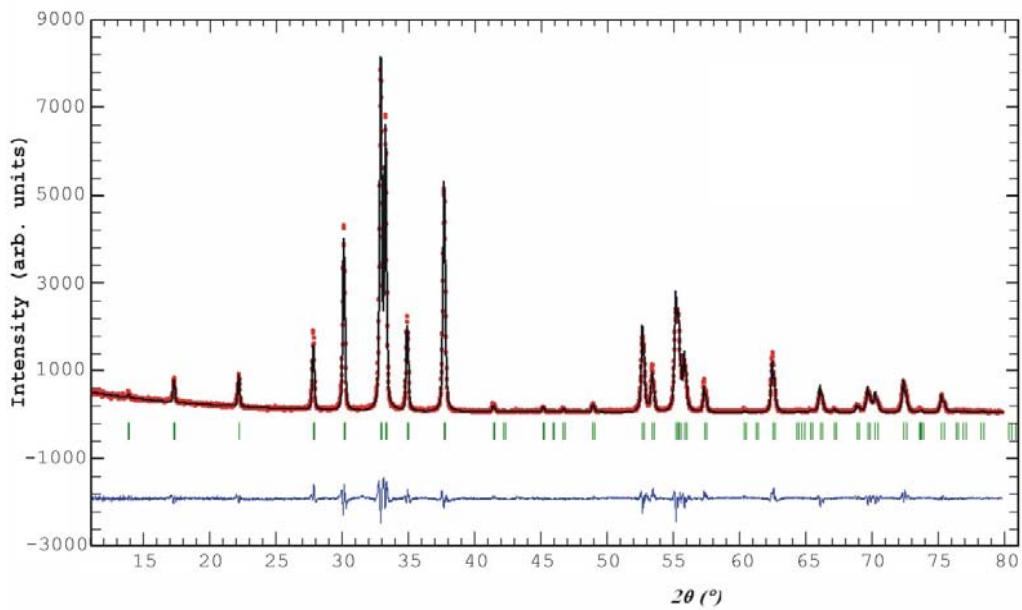

**Figure 5**

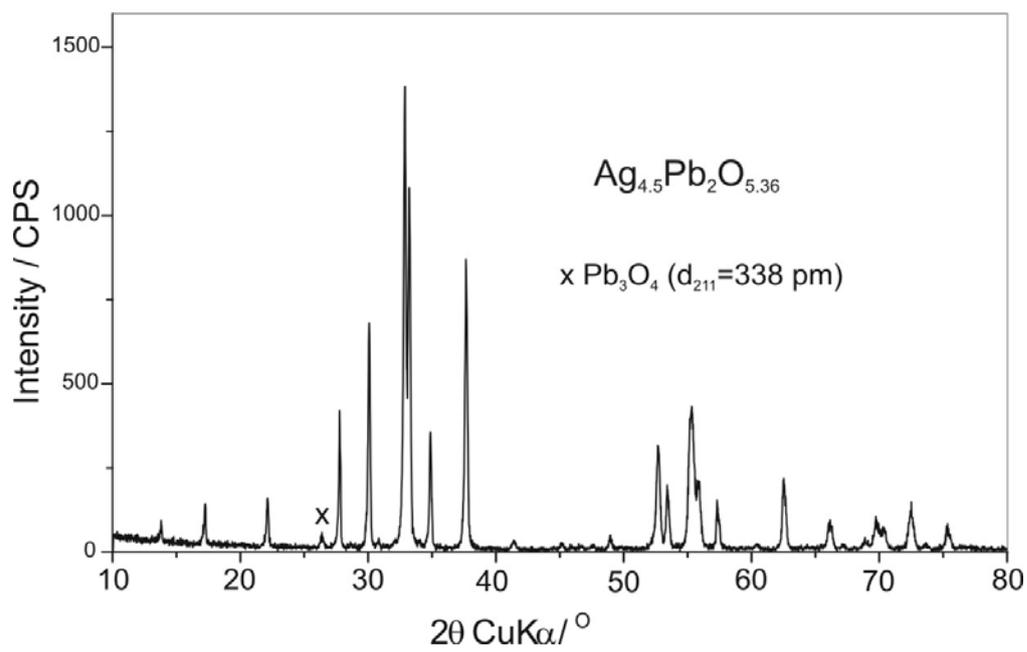

**Figure 6**

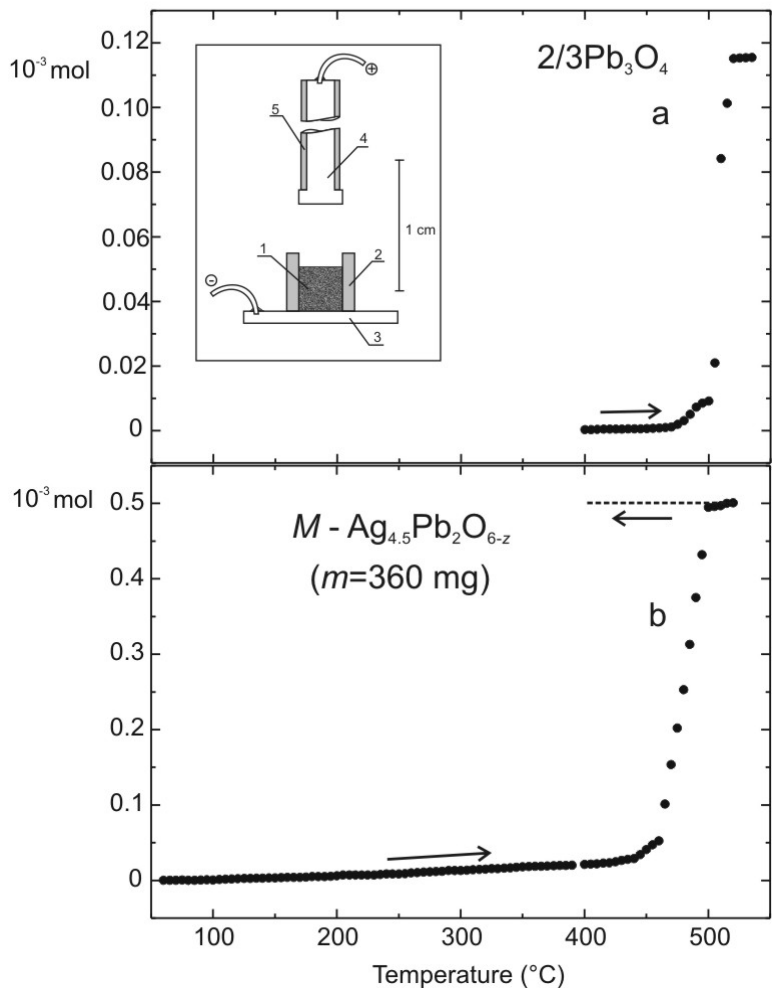

**Figure 7**

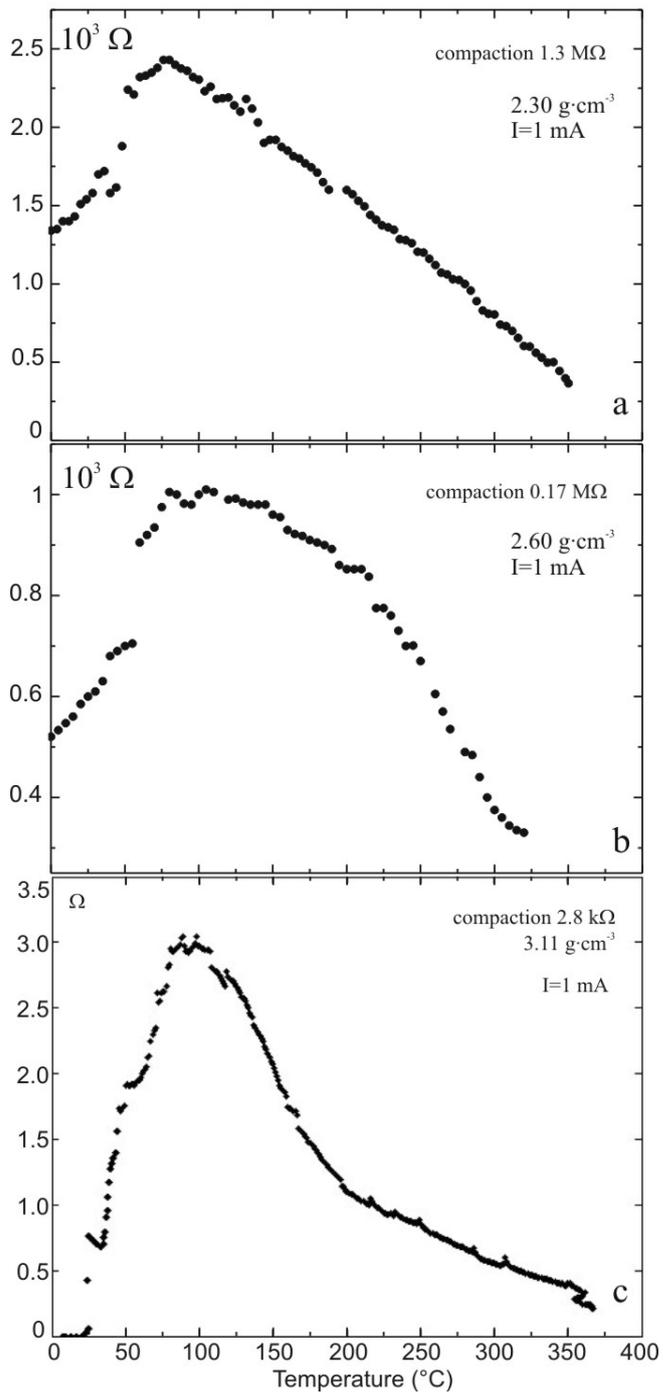

**Figure 8**

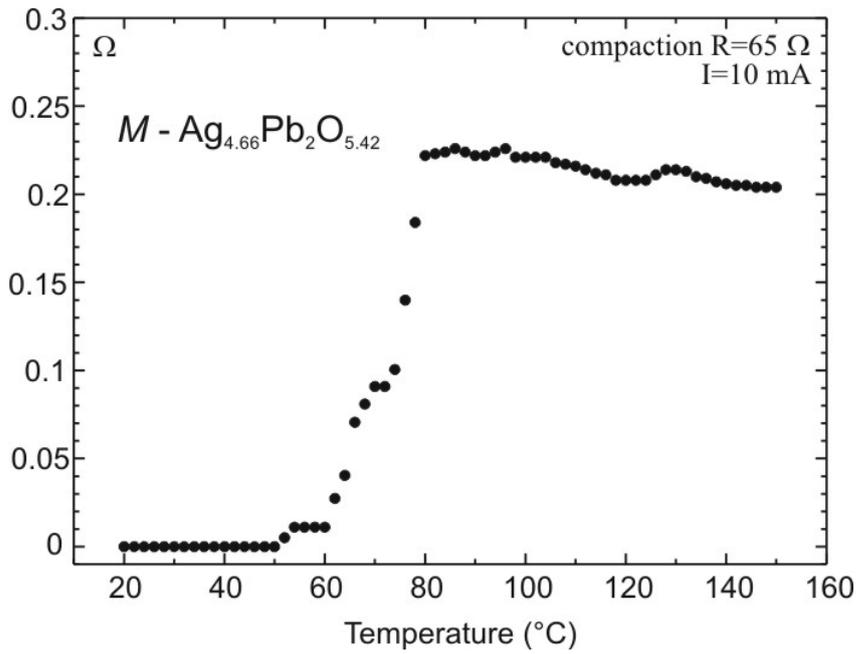

**Figure 9**

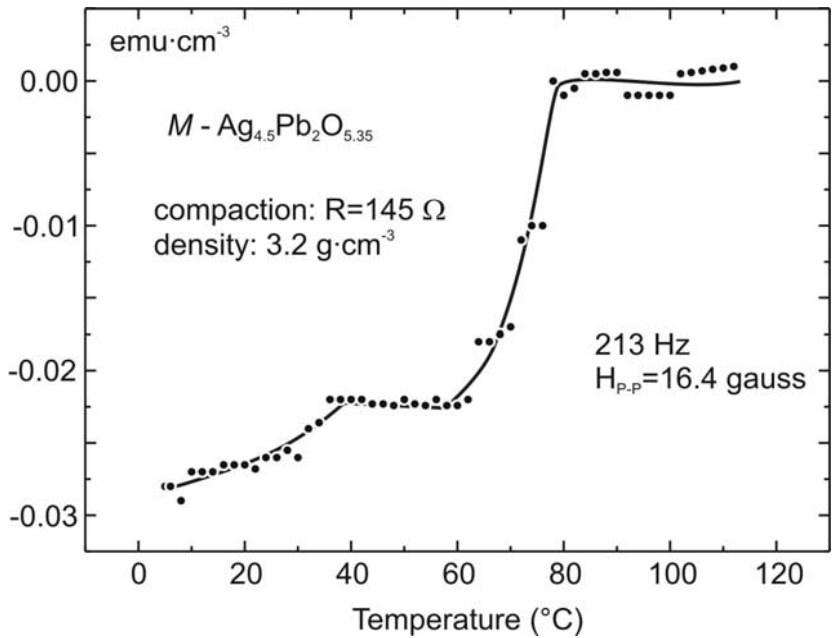

**Figure 10**